\documentclass[letterpaper,12pt]{article}
\usepackage{amsmath}
\usepackage{amsfonts}
\usepackage{amssymb}
\usepackage{graphicx}
\usepackage{physics}
\usepackage{latexsym}
\usepackage{epsfig}
\usepackage{pstricks}
\usepackage{stmaryrd}
\usepackage{rotating}
\usepackage[english]{babel}
\usepackage{setspace}
\usepackage[utf8]{inputenc}
\usepackage{natbib}
\usepackage[T1]{fontenc}
\usepackage{lmodern}
\usepackage{enumitem}
\usepackage{csquotes}
\usepackage{amsthm}
\usepackage{xcolor}
\begin{document}
%%%%%%%%%%%%%%%%%%%%%%%%%%%%%%%%%%%%%%%%%%%%%%%%%%%%%%%%%%
%%%%%%%%%%%%%%%%%%%%%%%%%%%%%%%%%%%%%%%%%%%%%%%%%%%%%%%%%%
\begin{center}	
\begin{LARGE}
\textbf{Assessing Relational Quantum Mechanics}\\
\end{LARGE}
\end{center}

\begin{center}
\begin{large}
R. Muciño, E. Okon and D. Sudarsky\\
\end{large}
\textit{Universidad Nacional Aut\'onoma de M\'exico, Mexico City, Mexico.}\\
\end{center}

Relational Quantum Mechanics (RQM) is an interpretation of quantum theory based on the idea of abolishing the notion of \emph{absolute} states of systems, in favor of states of systems \emph{relative} to other systems. Such a move is claimed to solve the conceptual problems of standard quantum mechanics. Moreover, RQM has been argued to account for all quantum correlations without invoking non-local effects and, in spite of embracing a fully relational stance, to successfully explain how different observers exchange information. In this work, we carry out a thorough assessment of RQM and its purported achievements. We find that it fails to address the conceptual problems of standard quantum mechanics---related to the lack of clarity in its ontology and the rules that govern its behavior---and that it leads to serious conceptual problems of its own. We also uncover as unwarranted the claims that RQM can correctly explain information exchange among observers, and that it accommodates all quantum correlations without invoking non-local influences. We conclude that RQM is unsuccessful in its attempt to provide a satisfactory understanding of the quantum world.

%\tableofcontents
\onehalfspacing
%%%%%%%%%%%%%%%%%%%%%%%%%%%%%%%%%%%%%%%%%%%%%%%%%%%%%%%%%%
%%%%%%%%%%%%%%%%%%%%%%%%%%%%%%%%%%%%%%%%%%%%%%%%%%%%%%%%%%
\section{Introduction}
\label{Intro}
%%%%%%%%%%%%%%%%%%%%%%%%%%%%%%%%%%%%%%%%%%%%%%%%%%%%%%%%%%
%%%%%%%%%%%%%%%%%%%%%%%%%%%%%%%%%%%%%%%%%%%%%%%%%%%%%%%%%%

In \cite{Rov96}, an interpretation of quantum theory, called \emph{Relational Quantum Mechanics} (RQM), is introduced. The interpretation is built around the proposal to get rid of the notion of ``\emph{absolute} states of systems'' in favor of ``states of systems \emph{in relation to each other}'' (see also \cite{Rov97,Rov97a,Rov07,Rov08,Rov18}). Such a move is argued to solve most, if not all, of the conceptual problems of standard quantum mechanics. Moreover, in spite of its fundamentally relational character, RQM has been claimed to contain enough resources to fully describe nature and to adequately explain the possibility of exchanging information about the world between observers. RQM has also been argued to account for all quantum correlations, without the need of invoking mysterious non-local influences.

The objective of this work is to carry out a thorough assessment of RQM (other works exploring RQM include \cite{Laudisa00,Norsen07,Brown09,Dieks09,van10,Wood10,Dorato16,Candiotto17,Ruyant18,Laudisa19,Pienaar19,Calosi20}). In particular, we first explore whether RQM is in fact able to solve the conceptual problems of standard quantum mechanics---and whether, while trying to do so, it stays clear of creating conceptual problems of its own. Then, we examine the claims that RQM is able to explain the possibility of exchanging information between observers, and that it fully foregoes mysterious non-local influences.

To do this, we begin in section \ref{Over} with an overview of RQM. Then, in section \ref{Assess}, we divide our assessment in four subsections: measurement, ontology, consistency and locality. In section \ref{Conc} we state our conclusions.

%%%%%%%%%%%%%%%%%%%%%%%%%%%%%%%%%%%%%%%%%%%%%%%%%%%%%%%%%%
%%%%%%%%%%%%%%%%%%%%%%%%%%%%%%%%%%%%%%%%%%%%%%%%%%%%%%%%%%
\section{Overview of Relational Quantum Mechanics}
\label{Over}
%%%%%%%%%%%%%%%%%%%%%%%%%%%%%%%%%%%%%%%%%%%%%%%%%%%%%%%%%%
%%%%%%%%%%%%%%%%%%%%%%%%%%%%%%%%%%%%%%%%%%%%%%%%%%%%%%%%%%

RQM is a non-standard, realist interpretation of quantum theory, first proposed in \cite{Rov96} with quantum gravity as a remote motivation. Since then, the theory has been developed and interests in it has grown steadily. RQM is built around the idea of abolishing the notion of absolute states of systems, and substituting it with states of systems relative to each other. Moreover, according to RQM, physical variables take values only at interactions.

In contrast with the standard interpretation, in RQM the role of the observer can be played by any physical system. Moreover, any interaction counts as a measurement. As a result, RQM is argued to solve the conceptual problems of standard quantum mechanics, including the (in)famous measurement problem. In particular, according to its proponents, RQM is able to make sense of the quantum world without requiring hidden variables, many worlds, physical collapses, or a special role for observers.

The points of departure in Rovelli's construction of RQM are: i) the realization that the standard interpretation of quantum mechanics is conceptually problematic, and ii) the hope that the problems can be solved, not by replacing or fixing the theory, but by truly understanding what it says about the world.

According to Rovelli, the core conceptual difficulty of the standard interpretation consists of the fact that the theory implies that a quantity can be both determined and not determined at the same time. To show this, he considers a Wigner's friend scenario, with observer $O$ measuring system $S$ inside a sealed lab, and observer $P$ standing outside. He then reasons as follows. Suppose that $O$ measures quantity $q$, which can only take values 1 and 2. According to the standard interpretation, if initially $S$ is prepared in a superposition of $q=1$ and $q=2$, then, after $O$ measures, he\footnote{As in \cite{Rov96}, we refer to $O$ as ``he'' and to $P$ as ``she.''} will find either $q=1$ or $q=2$ (with probabilities dictated by Born's rule). $P$, on the other hand, does not interact with the $S-O$ system. Accordingly, the standard interpretation asserts that, from $P$'s point of view, the $S-O$ system will evolve linearly, ending up in a superposition of $O$ measuring $q=1$ and $O$ measuring $q=2$.

Based on these observations, Rovelli argued that the standard interpretation \emph{forces} us to accept that ``different observers may give different accounts of the same sequence of events''. He took such an assertion as the basis for the construction of RQM (\cite{Rov96}):
\begin{quotation}
If different observers give different accounts of the same sequence of events, then each quantum mechanical description has to be understood as relative to a particular observer. Thus, a quantum mechanical description of a certain system (state and/or values of physical quantities) cannot be taken as an ``absolute'' (observer-independent) description of reality, but rather as a formalization, or codification, of properties of a system \emph{relative} to a given observer.
\end{quotation}
That is, Rovelli proposes to discard the notions of absolute state of a system, or value of a quantity, and to replace them by states or values \emph{relative} to different observers. 

On top of this, Rovelli postulates no system to play a privileged role. In particular, he assumes that there are no special systems, e.g., observers or measurement apparatuses, for which quantum laws do not apply. As a result, for him the word ``observer'' represents any sort of system, without it having to be conscious, animate, macroscopic, or special in any other way. He also takes quantum mechanics to be complete. Putting everything together, Rovelli finally concludes that \emph{what quantum mechanics does is to provide a complete description of physical systems, relative to each other}. Still, \emph{he takes such a description to exhaust all there is to say about the world}. These, in short, are the central tenets of RQM. 

There are a number of further details about RQM that we need to mention before moving on to our general assessment. First, we comment on RQM's stance regarding value assignment during measurements; that is, on how and why, say, $O$ (but not $P$), obtains a well-defined value of $q$ when he measures. According to RQM, since $P$ does not interact with the $S-O$ system, then she describes the measurement as a fully unitary process (leading to the establishment of correlations between $S$ and $O$, but not a definite value for $q$). $O$, on the other hand, only describes system $S$, but since the evolution of $S$ is affected by its interaction with $O$, the unitary description given by $O$ breaks down, and a defined value is acquired.

According to Rovelli, this breakdown of unitarity is not brought about by ``mysterious physical quantum jumps'' or ``unknown effects''. It arises instead because, for him, \emph{it is impossible to give a full description of an interaction in which one is involved}. As a result, $O$ cannot fully describe its interaction with $S$, so the unitary description breaks down and a concrete value of $q$ is produced for $O$. Of course, $O$ could be included in a larger $S-O$ system (remember that $O$ can be any ordinary system, for which quantum laws certainly apply), but then we would need to consider a description of such a system by a different observer (e.g., $P$) for which no definite value of $q$ would be obtained. It is important to point out that, on top of the previous explanation, in the recent \cite{Rov20} it is further argued that \emph{decoherence} plays a crucial role in explaining the breakdown of unitarity.

From what we have said so far, it can be concluded that, within RQM, value acquisition happens whenever \emph{any} pair of systems undergoes \emph{any} sort of mutual interaction---of course, we must not forget that this value acquisition happens only relative to the systems involved. We see, then, that RQM proposes some sort of democratization of the standard interpretation---which, of course, holds that definite results are obtained only when special systems, i.e., ``observers'' are involved in special interactions, i.e., ``measurements''. From this point of view, and as hoped by Rovelli, RQM is seen more as a generalization of the standard interpretation than a replacement or modification thereof.

As we just saw, according to RQM, value acquisition occurs when interactions happen. That raises the question as to exactly \emph{when} do those ``actualizations'' occur. The answer given by Rovelli is that they do so at the time at which the interaction takes place and, as argued in \cite{Rov98}, RQM allows for a clear-cut definition of such a time. More precisely, Rovelli claims that RQM allows for the definition of a probability distribution for the time of the interaction in terms of the probability distribution for the establishment of a correlation between the systems involved, as measurable by a third observer. Going back to the Wigner's friend scenario, the idea is that $P$ can define a probability distribution for the time of $O$'s measurement, as the expectation value of an operator $M$, which has as eigenstates those states in which $O$ successfully measures the value of $q$.

Given its relational character, a potential danger for RQM is for it to lead to inconsistencies when different descriptions of the same system or situation are considered. For instance, couldn't $P$ and $O$ get different answers out of measurements of the same system $S$? Rovelli warns that such a question must be handled with care because, within RQM, the information possessed by different observers cannot be compared freely. Two observers can indeed compare information, but they can do so only by physically interacting with each other. What is meaningful to ask within RQM, it is then argued, is whether $P$'s measurement of $S$ coincides, not with what $O$ found when he measured $S$, but with what $O$ tells $P$ that he found when $P$ asks $O$ about his result. According to Rovelli, quantum mechanics automatically guarantees this kind of consistency because, when $P$ measures $S$, she collapses the state of the full $S-O$ system, ensuring that the second measurement (asking $O$ what he got) would yield a consistent answer. Finally, Rovelli asserts that this guaranteed consistency clears out all clouds of solipsism haunting RQM.

Next, we comment on the ontology of RQM, i.e., on what (if anything) the theory takes as actual ``elements of reality''. First, we note that Rovelli is emphatic that, in RQM, the wave function is not real; it only codifies relational information, in the form of correlations between different systems, which can be used to make future predictions. Still, Rovelli argues that RQM can, and should, be given a realist reading. What constitutes physical reality in RQM, Rovelli argues, are what he calls ``quantum events'', which are given by the actualization of definite values of properties that occur in the course of interactions. Given that these actualizations are relative and discrete, so are Rovelli's quantum events. Accordingly, the world of RQM has a sparse ontology constructed by an observer-dependent ensemble of discrete quantum events. Nevertheless, in \cite{Rov20} it is argued that, out of this sea of relative events, decoherence causes the emergence of \emph{stable} events or facts, whose relativity can effectively be ignored---thus explaining the emergence of the classical world.

Lastly, we briefly review what has been said along the years regarding the status of locality within RQM. In \cite{Rov07}, it is argued that if one accepts that, in the absence of a physical interaction between observers, it is meaningless to compare measurements occurring at space-like separation, then the EPR correlations do not entail any form of non-locality. From that, they conclude that RQM is a fully local theory. However, in \cite{Rov19}, such an assertion is walked back and it is conceded that RQM is in fact non-local in the sense of Bell. Still, it is argued that, within RQM, such a failure of locality is not deep or interesting, since it simply ``reduces to the existence of a common cause in an indeterministic context''.

In sum, RQM is a \emph{realist} interpretation of quantum theory built around: i) the eradication of absolute states of systems, in favor of states of systems relative to each other and ii) proposing that physical variables take values only at interactions. These moves are claimed to solve the conceptual problems of standard quantum mechanics. Moreover, RQM is said to accommodate all quantum correlations without non-locality and, in spite of its relational character, to be able to explain the interchange of information between agents.

This concludes our overview of RQM. In the next section we explore these ideas through a much more critical lens. 

%%%%%%%%%%%%%%%%%%%%%%%%%%%%%%%%%%%%%%%%%%%%%%%%%%%%%%%%%%
%%%%%%%%%%%%%%%%%%%%%%%%%%%%%%%%%%%%%%%%%%%%%%%%%%%%%%%%%%
\section{Assessment of Relational Quantum Mechanics}
\label{Assess}
%%%%%%%%%%%%%%%%%%%%%%%%%%%%%%%%%%%%%%%%%%%%%%%%%%%%%%%%%%
%%%%%%%%%%%%%%%%%%%%%%%%%%%%%%%%%%%%%%%%%%%%%%%%%%%%%%%%%%

After this review of RQM, we are ready to explore in detail the interpretation. We start by examining whether RQM is able to solve the conceptual problems of standard quantum mechanics, and whether, in its attempt, it stays clear of introducing fresh problems of its own. Next, we explore whether RQM is capable to fully describe nature and to accommodate the possibility of exchanging information between observers. Finally, we analyze the claim that RQM fully foregoes mysterious non-local influences. 

%%%%%%%%%%%%%%%%%%%%%%%%%%%% %%%%
\subsection{Measurement}
\label{M}
%%%%%%%%%%%%%%%%%%%%%%%%%%%%%%%%

As we explained above, the main motivation behind RQM is to come up with an interpretation of quantum mechanics that addresses the conceptual problems of the standard framework (e.g., the measurement problem). In order to evaluate the success of RQM in this regard, we start by examining in some detail these conceptual problems.

As is well-known, the standard, linear, quantum evolution, when applied to macroscopic objects, generically leads to superpositions of macroscopically different states. In order to avoid this undesirable result, standard quantum mechanics contains two very different evolution laws: i) the linear and deterministic Schrödinger equation and ii) the non-linear and indeterministic collapse process. The dynamics is postulated to be generically governed by the Schrödinger equation, but when a \emph{measurement} takes place, the collapse process is assumed to come into play, interrupting the Schrödinger evolution. As a result, the formalism ends up critically depending upon the notion of measurement---which is a problem because such a notion is never precisely defined within the theory. And it is not only that the standard theory does not specify when a measurement happens, it also does not prescribe what it is that is being measured (i.e., in which basis will the collapse occur). The upshot is a formalism with an unacceptable vagueness at its core. This, in brief, is what is usually referred to as \textit{the measurement problem} (see \citep[section 4]{Myrvold} for an overview).

One way to state the problem, as Bell argued in \cite{Bel:90}, is that higher-level notions, such as measurements, should not appear as primitives in a theory that aims at being fundamental. Therefore, to solve the problem, a physical theory must be fully formulated in mathematical terms. As a result, concepts such as measurement, measuring apparatus, observer or macroscopic should not be part of the fundamental language of the theory. Along this, one must make sure that successful applications of the formalism do not require the introduction of information that is not already contained in the fundamental description of the theory for the situation under consideration. That is, one must make sure that, once a complete quantum description of a physical scenario is given---including a full description of the quantum state of the measuring apparatus and all the details of its interaction with the rest of the system---then, with that information alone, it is possible to make concrete predictions regarding what are the possible final outcomes of the experiment.

To illustrate this particular face of the measurement problem (also known as the \textit{preferred basis problem}), consider a particle and a pair of two-level detectors (with levels $|u\rangle$, unexcited, and $|e\rangle$, excited) located at $x = D$ and $x = -D$, respectively. Initially, the detectors are unexcited and the particle's wave function is a narrow Gaussian centered at $x = 0$, that is
\begin{equation}
	\Psi (0) =| \psi_{0} \rangle \otimes | u \rangle_+ \otimes | u \rangle_- .
	\end{equation}

As for the Hamiltonians, the free Hamiltonian of the particle is
\begin{equation}
		\hat{H}_p^{0} = \frac{\hat{p}^2}{2m}
	\end{equation}
and those of the detectors are
	\begin{equation}
		\hat H_\pm^{0}= \epsilon \lbrace | e \rangle_\pm \langle e |_\pm - | u \rangle_\pm \langle u |_\pm \rbrace.
		%\right],
	\end{equation}
Finally, the Hamiltonians describing the interaction of the particle and detectors is 
	\begin{equation}
		\hat H_\pm^{I} = \lambda \delta (\hat x \mp D \hat{I_p}) \otimes \left(| e \rangle_\pm \langle u |_\pm + | u \rangle_\pm \langle e |_\pm \right) .
	\end{equation}

Given all this, it is easy to see that, after some time $t$, the state will evolve to
\begin{eqnarray}
	\Psi (t) & = &
	| \psi_{eu} \rangle\otimes | e \rangle_+ \otimes | u \rangle_- +| \psi_{ue} \rangle \otimes | u \rangle_+ \otimes | e \rangle_- \nonumber \\
	& & + | \psi_{uu} \rangle \otimes | u \rangle_+ \otimes | u \rangle_- + | \psi_{ee} \rangle \otimes | e \rangle_+ \otimes | e \rangle_- .
\end{eqnarray}
It would seem that one can interpret each of the terms above easily and make the corresponding predictions: either one or the other of the detectors is excited, no detection has taken place (because no detector is perfect) or there has been a double detection (involving something like a bounce, and corresponding to a small effect of order $\lambda^2$). However, the point we want to make is that, one might consider, instead, describing things in terms of an alternative basis for the detectors pair. For instance,
\begin{equation}
	| E \rangle \equiv | e \rangle_+ \otimes | e \rangle_- \nonumber
\end{equation}
\begin{equation}
	| U \rangle \equiv | u \rangle_+ \otimes | u \rangle_- \nonumber
\end{equation}
\begin{equation}
	| S \rangle \equiv \frac{1}{\sqrt 2 } \left[ | e \rangle_+ \otimes | u \rangle_- + | u \rangle_+ \otimes | e \rangle_- \right] \nonumber
\end{equation}
\begin{equation}
	| A \rangle \equiv \frac{1}{\sqrt 2 } \left[ | e \rangle_+ \otimes | u \rangle_- - | u \rangle_+ \otimes | e \rangle_- \right] .
\end{equation}
Now, if the detectors are macroscopic, we know that the experiment will end up in any of the terms of the original basis, and not in any of the last two terms above---such states do not correspond to macroscopic objects in well-localized states. The key question, of course, is whether standard quantum mechanics is capable of \emph{predicting} this fact. The answer is that it is not. We know from experience that macroscopic objects always possess well-defined positions, the problem is that it is impossible to derive such a result from the standard formalism alone, even when, as above, a full quantum description of the whole system is provided. That is, without information not contained in a complete quantum description, standard quantum mechanics is unable to deliver concrete predictions regarding the possible final outcomes of an experiment.%\footnote{In practice, such additional information is supplied by the implicit assumption, arrived at by experience, that macroscopic objects, such as measuring devices and observers, always possess well-defined positions---a fact which, as we explained, is not dictated by the theory itself.} 

It has been argued that \emph{decoherence} is able to resolve this issue. In particular, it has been argued that decoherence is able to explain how, by taking into account the interaction between systems and their environment, a preferred basis \emph{dynamically} emerges. There are, however, problems with such an explanation. To begin with, the division of the world into a system and an environment is totally arbitrary. As a result, different decisions as to how to split a system will lead to different preferred bases. Moreover, even if decoherence were able to select a basis, given its inability to resolve the measurement problem, it completely lacks the resources to link such a basis with what we actually expect to see in an actual experiment (for details, see \cite[section 2]{Less}).

It is important to keep in mind that the conceptual problems of standard quantum mechanics do not end with the above mentioned ambiguities regarding the dynamics. The framework is also problematic as a realist physical theory because it lacks ontological clarity (see \cite{maudlin2018ontological} for a illuminating discussion about this last notion). That is, the framework does not allow for it to be interpreted as providing a description of the world it is supposed to portray. This might be acceptable if all one wants is a tool to make predictions in suitable experimental settings, but is untenable if one aims at constructing a complete, fundamental physical theory. It is true that, by adopting an empiricist or a pragmatist position (see, e.g., \cite{vanFraassen2001} or \cite{Healey2012QuantumTA}), the requirement of ontological clarity is relaxed. However, Rovelli explicitly intends RQM to be given a realist reading, so the demand for ontological clarity remains.\footnote{One could argue that the stance on the conceptual problems of quantum mechanics described in the last pages is only a point of view, as valid as many others. However, a demand of clarity, precision or self-consistency in a realist physical theory is not a negotiable ``point of view'' that one can simply decide not to adhere to.}

In this section, we explore how RQM fares as a solution to the problems mentioned in the previous paragraphs; in the next, we return to these ontological issues and analyze them in connection with RQM.

As we explained above, for RQM, the key observation regarding the conceptual content of quantum mechanics is the fact that the theory allows for a quantity to be both determined and not determined at the same time.\footnote{In fact, in \cite{Rov96} it is claimed that ``the experimental evidence at the basis of quantum mechanics forces us to accept that distinct observers give different descriptions of the same events''; that ``it is nature itself that is forcing us to this way of thinking''. But that, of course, is simply not true, as there are quantum formalisms, such as pilot-wave theory (\cite{Bohm}) or objective collapse models (\cite{GRW,CSL,Bassi}), that accommodate all the empirical evidence at the basis of quantum mechanics as well than the standard framework, and do so without ever allowing for distinct observers to give different descriptions of the same events. In any case, in a recent private communication, Rovelli informed us that his point of view is now different, namely that ``if you take textbook quantum mechanics at its face value, it forces you to this conclusion.''} However, it is the vagueness of the collapse postulate---the ambiguity as to what causes it and what it entails---that allows for the discrepancy between descriptions. That is, a core conceptual problem of standard quantum mechanics has to do with an ambiguity regarding the dynamics. Such an ambiguity allows for different observers to give different descriptions of the same events. RQM takes this fact to be the underlying cause of the problem and constructs a formalism that removes the tension associated with the discrepancy. The problem, as we will see below in detail, is that such a strategy fails in addressing the true, underlying conceptual problems.

We start by exploring in more detail RQM's stance regarding value acquisition during measurements. As we saw in section \ref{Over}, in the context of a Wigner's friend scenario, RQM's explanation on why $P$ does not obtain a well-defined value of $q$, but $S$ does, runs as follows. Given that $P$ does not interact with the $S-O$ system, she describes the measurement of $S$ by $O$ as a purely unitary process, which leads to a correlation between $S$ and $O$, but not to a definite value for $q$. $O$, on the other hand, attempts to assign to system $S$ a unitary description, but since the measurement interaction affects the evolution of $S$, $O$'s unitary description of $S$ breaks down, and a defined value is obtained.

The important point, though, is that, for Rovelli, this breakdown of unitarity is not due to an obscure physical jump or collapse; it is instead a consequence of the fact that $O$ is unable to give a full description of his interaction with $S$. This, in turn, is supposed to follow from the fact that such a description would contain information regarding correlations involving $O$ himself but, it is argued, ``there is no meaning in being correlated with oneself'' \cite{Rov96}. 

This last claim seems to play a crucial role in the whole account, but where does it come from? In order to support it, besides invoking his intuitions, Rovelli alludes to a theorem proven in \cite{Breuer}, which shows that ``it is impossible for an observer to distinguish all present states of a system in which he or she is contained.'' If so, it is argued, it is impossible for an observer to give a full description of an interaction in which she is involved. 

Regarding the theorem, what it actually shows is that, if you assume that a measuring apparatus is part of a system, then the states of the apparatus cannot fully codify the state of the whole system (including the apparatus itself). That is, it shows that, by looking only at the state of the apparatus, one cannot fully infer the state of the whole system. It is not clear, however, why would one expect a measuring apparatus to be able to do that. For instance, even if the measuring apparatus has enough degrees of freedom to codify all the remaining degrees of freedom of the system, just by looking at the apparatus, it is impossible to know if the degrees of freedom of the apparatus are in fact correlated with those outside of it. We see that the theorem depends upon a highly contentious requirement, so it is not clear what the real-life consequences of the theorem actually are.

What about the use of the theorem in order to supports RQM's claim that there is no meaning of the correlations of a system with itself? As we said, the interpretation of the theorem is debatable. At any rate, we must say that, regardless of how it is interpreted, we find the connection between the theorem and the question of how quantum values are acquired quite obscure. The point is that, even if the theorem in fact shows that an observer cannot fully describe a system which includes herself, it is not at all clear how that is related to a change in the physical evolution of the system in question. That is, it seems that the claim is that an epistemic limitation would have a profound physical implication. Moreover, it is not at all clear, even if one were to accept the argument (and we certainly do not) that a breakdown of the description via the unitary evolution should occur, why would that breakdown take the particular form it takes. That is, why would it consist of properties of $ S$ acquiring, for $ O$, definite values corresponding to eigenvalues of particular operators, and why would the probabilities of these occurrences be given by the Born rule. Rovelli claims that there is nothing mysterious involved in the breakdown of unitary, but the explanation he offers seems to us to be quite mysterious.

Returning to the statement that there is no meaning in being correlated with oneself, we must say that we find such a claim quite odd. To begin with, it is clear that the different parts of an observer certainly are correlated between them: her left hand is never more that 2 meters away from the right one. Moreover, even considering an observer as a whole, we find it much more reasonable, not only to say that it can, of course, be correlated with herself, but that the correlation is, in fact, always perfect: given her state, one can certainly predict with certainty her state. At any rate, the truth of the matter is that the whole discussion regarding the impossibility of a system having correlations with itself is a red herring. To see this, we remember that what needs explaining is the transition from a superposition in which $O$ measures both $q=1$ and $q=2$ to a state in which he measures either $q=1$ or $q=2$. The important point, though, is that what distinguishes the situation before and after the actualization of the value of $q$ is the disappearance of the superposition, not of correlations. That is, when the actualization occurs, a correlation between $S$ and $O$ certainly prevails. It is clear, then, that focus on correlations and, in particular, on which correlations can or cannot occur, is irrelevant in order to explain the phenomenon of value acquisition during measurements.

As we said above, recent papers on RQM have omitted the explanation of value actualization described above and have argued that decoherence plays a crucial role in explaining the breakdown of unitarity (e.g., \cite{Rov20}). This claim, of course, is nothing new (see e.g. \cite{Zur,Sch}). Also not new is a clear understanding of the fact that decoherence alone is not able to explain the apparent breakdown of unitarity. \cite{Rov20} acknowledges that, but it is claimed that this is because decoherence, by itself, lacks an ontology. Moreover, it is argued that the ontology provided by RQM solves the issue. That is, that the combination of decoherence and RQM does provide a sound explanation for the breakdown of unitarity. The problem is that the limitations of decoherence as an explanation of the collapse process go much deeper than not providing a clear ontology. The fact is that decoherence fails in this regard because, in order for the explanation it provides to actually work, one basically needs to assume what one wants to explain. One can see why this is so by recalling that the explanation for the breakdown of unitarity provided by decoherence involves the claim that, \emph{for all practical purposes}, a decohered reduced density matrix behaves as a (proper) mixture. The problem is that this is so only if one assumes that, when one measures, one always finds as a result an eigenstate of the measured operator. In other words, in order to explain the breakdown of unitarity through decoherence, one needs to precisely assume a breakdown of unitarity. That is, in order to show that the reduced density matrix of a system behaves, for all practical purposes, as the corresponding mixed density matrix, one needs to assume that, upon a measurement, systems collapse according to the collapse postulate and the Born rule (for a detailed explanation of why this is the case, see \cite[section 2]{Less}). And, clearly, focusing on relative states does nothing to alleviate the issue. We conclude that RQM does not provide to decoherence what it lacks in order to solve the problem.

Independently of the details of the collapse process within RQM discussed above, what is clear is that, according to the formalism, value acquisition occurs when \emph{interactions} happen. It seems, then, that the framework needs for there to be a well-defined moment at which each interaction takes place; otherwise, the proposal becomes vague and loses all strength. To address the issue, \cite{Rov98} argues that RQM allows for a clear-cut definition of such a time. The proposal is that the time of an interaction is only defined probabilistically, and relative to an outside observer. More concretely, RQM proposes to define a probability distribution for the time of the interaction in terms of the probability distribution for the establishment of a correlation between the systems involved, as measured by an outside observer. For instance, if we consider $O$'s measurement of $S$ in the Wigner's friend scenario, the initial state is a separable state in which $O$ is ready to measure and $S$ is in a superposition of $q=1$ and $q=2$. Such a state, via purely unitary evolution, transforms into a superposition of $O$ measuring both $q=1$ and $q=2$. The idea, then, is that $P$ can define a probability distribution for the time of the measurement as the expectation value of an operator, $M$, that has as eigenstates those states in which $O$ successfully measures the value of $q$. The idea, in sum, is that measurements happen when certain correlations are established.

We find the proposal problematic for various reasons. To begin with, we find it strange that the time of a measurement is defined, not for the observer for which, within RQM, such a measurement is actual and leads to a well-defined value, but for outside observers for whom, according to RQM, such an interaction cannot be said to have happened. Moreover, different outside observers will assign different values for the time of the interaction. In \cite{Rov98} it is pointed out that all this is not inconsistent, and we agree; but not being inconsistent is not enough to make the proposal reasonable.

In any case, the more serious complication stems from the fact that, when examined in full detail, the proposal reveals actually not to be well-defined. The problem is that the proposed definition explicitly depends on the operator $M$, but such an operator is, to a large extent, arbitrary. This, is because it depends on a \emph{convention} as to which states of a measuring apparatus codify or stand for which states of the measured system. Without that information, not contained in the quantum description, there is no way to define $M$.\footnote{Of course, all this is a consequence of the more general ambiguity problem that, we noted, affects standard quantum theory when deprived of special roles for measuring devices or observers. If anything, this proposal makes this problem more apparent, by noticing the need for a choice of the operator $ M$, but failing to recognize that such a choice is in no way dictated by the postulates of the theory.}

To see this, consider a free spin-$\frac{1}{2}$ particle to be measured by an apparatus containing a macroscopic pointer, whose center of mass $y$ has an initial state given by a narrow wavefunction $\varphi (y)$ centered at $y=0$. Assume, for simplicity, that the free Hamiltonians of the pointer and the spin are zero, and that the spin and the apparatus interact via
\begin{equation}
\hat{H}_{I}=2 i \hbar \lambda \left( \frac{\partial}{\partial y} \right) \otimes S_z
\end{equation}
with $\lambda$ a constant. Now, if the initial state of the spin is up along $x$, then the state of the complete system at time $t$ will be given by
\begin{equation}
\ket {\Psi(t)} = \frac{1}{\sqrt{2}} \left [ \varphi(y-\lambda t) \otimes \ket {+_z}+ \varphi(y+\lambda t) \otimes \ket {-_z} \right ] .
\label{sup}
\end{equation}
Therefore, if $\lambda t$ is big enough, one can infer the value of the spin along $z$ by looking at the position of the pointer.

Going back to the issue of determining the time of the interaction, what is $M$ in that case? The issue is that the quantum description of the situation does not allow for such a question to be answered. In particular, it does not contain the information as to how much the pointer must move, in order for its position to correspond to a certain value of spin. In a concrete experimental situation, one could try to give a \emph{pragmatic} answer, in terms of how the actual measuring apparatus available is employed in practice. The problems is that if one takes quantum mechanics to be fundamental and complete, as RQM does, then all the relevant information to do physics must be contained in the quantum description itself. Moreover, in the context of RQM, one, of course, cannot employ this pragmatic attitude, as the definition should work for all sorts of systems and situations, not only those containing actual observers in standard laboratory settings.

Even worse, as in the discussion about the breakdown of unitarity above, the focus on correlations can be seen to be a distraction. This is because, what characterizes the transition from the initial state to the final one, is not the appearance of correlations but the appearance of entanglement. That is, the initial state is as correlated as any other: it is a separable state that correlates the initial ready state of $O$ with the superposition of $S$. Could focusing on the presence of entanglement solve the issue? Not really because, even if the initial state is separable, if an interaction Hamiltonian is present, entanglement would arise intermediately. Therefore, if one does not demand some quantitative level of entanglement, one would have to accept that there is always at least some level of it resulting from, say, the gravitational interaction. We conclude that RQM cannot really provide a reasonable definition for the time at which an interaction happens.

As we explained at the beginning of this section, in order to solve the conceptual problems of quantum theory it is necessary to remove the vagueness of the standard framework by constructing a framework fully written in precise, mathematical terms. In particular, the notion of measurement cannot be part of such a framework at the fundamental level. RQM tries to achieve this by proposing that interactions should play the role that measurements play in the standard framework. This might seem promising, as the notion of an interaction seems less observer-dependent than that of measurement. The problem is that, as we argued above, there is no well-defined procedure to define when interactions happen, which means that the proposal fails to resolve the problem. 

Moreover, above we also mentioned that having a precisely defined framework is not enough to solve the conceptual quantum problems. One also has to make sure that the constructed framework can in fact be employed to make predictions without requiring information not contained in the fundamental description given by the theory, for the scenario under consideration. The problem for RQM is that, even if, ignoring our criticisms, one were to grant that it achieves the former, it is unsuccessful regarding the latter.

To see this, consider again Wigner's friend scenario. The first step, which RQM takes to be unproblematic, involves $O$ measuring $S$ to obtain either $q=1$ or $q=2$. But already there there is a complication. The point is that, independently of the question of how a value for $q$ actualizes, there is the question of what determines the \emph{basis} in which such an actualization occurs. The fact is that if, as RQM demands, all is quantum, then there is simply no way to determine, out of the fundamental quantum description of the $S-O$ system (including the quantum description of the measuring apparatus, the full Hamiltonian, etc.), whether the actualization should occur in the $q=1$ or $q=2$ basis or in any linear combination thereof. That is, without some sort of additional information, not already contained in the (allegedly complete) quantum description, it is impossible to even predict what are the possible results of the experiment.

Regarding this so-called preferred basis problem, and the possibility of selecting one, \cite{Rov97a} states the following:
\begin{quotation}
...given an arbitrary state of the coupled $S-O$ system, there will always be a basis in each of the two Hilbert spaces which gives the bi-orthogonal decomposition, and therefore which defines an $M$ for which the coupled system is an eigenstate. But this is of null practical nor theoretical significance. We are interested in certain self-adjoint operators only, representing observables that we know how to measure; for this same reason, we are only interested in correlations between certain quantities: the ones we know how to measure.
\end{quotation}
It seems, then, that fully contingent, epistemic properties of the agents that navigate the world have a profound impact at the ontological level, i.e., such epistemic properties determine what there is in the world. We take this as a straightforward confession of the fact that RQM requires information not contained in the formalism to actually make predictions; that is, of the fact that RQM does not solve this conceptual problem of the standard interpretation.

To conclude this section, we display a serious complication with the basic notion of relative state, on which the whole RQM proposal is constructed. For this, consider the statement
\begin{quotation}
$\phi$: The state of the system $S$, relative to the system $O$, is $|\psi \rangle$.
\end{quotation}
Now, as stated in \cite{Rov96},
\begin{quotation}
A statement about the information possessed by $O$ is a statement about the physical state of $O$; the observer $O$ is a regular physical system. Since there is no absolute meaning to the state of a system, any statement regarding the state of $O$, including the information it possess, is to be referred to some other system observing $O$.
\end{quotation}
Moreover, the quantum description is taken to exhaust all there is to say about the world, which means that the quantum state should encode every relevant statement about the world, including the discourse of those who use it. Therefore, statement $\phi$ must correspond to (or, at least, be encoded in) a state, $|\chi \rangle $, of the system $S-O$, relative to some other system $P$. However, the previous statement, i.e., the claim that the state of $ S-O$, relative to $ P$, is $|\chi \rangle $, is a predicate about the system $S-O-P $, so it must correspond to (or, at least, be encoded in) a state of the system $ S-O-P $, relative to some other system $Q$... and so on. It seems, then, that in order for the basic statement $\phi$ to make any sense within RQM, it is necessary to invoke an infinite tower of observers! Moreover, such an infinite tower seems in serious danger of being ungrounded, which is known to lead to paradoxes (see \cite{Yablo1,Yablo2,Maudlin}).\footnote{Clearly, the necessity for an infinite tower of observers immediately disappears as soon as one allows for systems to possess absolute states.} 

In order to remedy this infinite tower issue, a couple of alternatives suggest themselves. However, both lead to further complications. The first option is to claim that $\phi$ is not a predicate about the $S-O$ system, but a predicate by $O$ about $S$, period. That is, that for each observer, there are determined values, without the need of a further observer. If so, it would be impossible for $O$ to describe, within RQM, its interactions with $S$ (or any other systems external to $O$). That is because, in order for $O$ to describe its interaction with system $S$, $O$ must be able to describe itself---but that is something that, in the context of RQM, $O$ cannot do. Given the emphasis that RQM often puts on the claim that acquisition of information necessarily involves physical interaction, this impossibility to describe interactions would seem to mean an impossibility to explain how one acquires information about the world. Moreover, all this would leave RQM in the same situation than standard quantum mechanics, in which, in order to describe a system, one necessarily requires an observer external to the studied system.

The other option would be to hold that $\phi$ is indeed a predicate about the $S-O$ system, but relative to $O$ itself. If so, it would not be necessary to invoke additional observers in order to make sense of $\phi$. However, this position seems inconsistent with the claim, employed by RQM to explain the breakdown of unitarity, that it is impossible to give a full description of an interaction in which one is involved. We conclude that the notion of a relative state, which is at the basis of the RQM construction, is extremely problematic.

To sum up, RQM promises to cure the conceptual problems of standard quantum mechanics. However, it fails to directly address such issues, and it leads to serious conceptual problems of its own. As we saw, an inherent, unavoidable vagueness in the dynamics is one of the most damaging conceptual problems of standard quantum mechanics. Such vagueness, in turn, allows for different observers to give different descriptions of the same sequence of events. RQM misreads this last feature for the true underlying cause and constructs a formalism to address it. The problem is that such a strategy fails to tackle the real issues.

For instance, within RQM, the breakdown of unitarity is not brought about by mysterious quantum jumps. Instead, it is a consequence of the fact that it is impossible to give a full description of an interaction in which one is involved. However, under scrutiny, such an explanation is found to be inadequate, as it focuses too much on the concept of a correlation, which turns out to be irrelevant. At the end, even though it is claimed that there is nothing mysterious involved in the breakdown of unitary, the explanation offered is exposed to be quite mysterious. \cite{Rov20} invokes decoherence as an explanation for the breakdown of unitarity, but fails to save decoherence from its well-known problems in this respect. We then explored a proposal to define the time at which an interaction happens, in terms of the establishment of correlations. However, since the notion of a correlation contains an arbitrary component, we found the definition lacking. Finally, we showed that RQM simply fails to address the so-called basis problem of quantum mechanics and that the notion of a relative state within RQM is problematic. We conclude that RQM fails to solve these conceptual problems of quantum mechanics.

%%%%%%%%%%%%%%%%%%%%%%%%%%%%%%%%
\subsection{Ontology}
\label{O}
%%%%%%%%%%%%%%%%%%%%%%%%%%%%%%%%

As we mentioned above, the conceptual problems of quantum mechanics also include the fact that the standard framework lacks \emph{ontological clarity}: that is, the fact that the formalism does not specify what kind of physical world it is supposed to describe. This is because, while it, of course, constitutes an extremely powerful predictive algorithm, it only makes predictions for the possession of properties, but remains fully silent regarding what are supposed to be the property bearers of the theory. This situation might be adequate if all one is interested in is a tool for making predictions in suitable experimental settings; however, it is unsatisfactory if one aims at constructing a proper physical theory, which specifies what exists in the world and how it behaves.

Is RQM able to cure this problem of the standard formalism? That is, unlike the standard framework, can RQM be read as providing a picture of the world it is supposed to portray? For this, we need to explore what, if anything, the theory takes to constitute physical reality. 

As we explained above, within RQM, the wave function is not real. Instead, it is taken to be a mathematical tool to codify information regarding correlations between different systems, which can then be used to make predictions. Still, Rovelli insists that RQM should be given a realist reading. To do so, he proposes to take as fundamental elements of physical reality, so-called ``quantum events'', which are given by the actualization of definite values of properties that occur in the course of interactions. Given that, according to RQM, such actualizations are relative and discrete, so are these quantum events. As a result, the world of RQM is argued to be constructed by an observer-dependent, sparse, collection of discrete quantum events.

As an example of a quantum event, the detection of an electron in a certain position is offered. The idea is that ``the position variable of the electron assumes a determined value in the course of the interaction between the electron and an external system and the quantum event is the `manifestation' of the electron in a certain position'' \cite{Rov08}. 

The first thing we note about the proposal is that the example provided could be quite misleading. This is because, while the quantum event in such an example could be thought of as a point in space-time,\footnote{\cite{Rov19} explicitly states that the `elements of reality' must live in space-time.} and thus as a potential candidate to constitute the ontology of the theory, this is not the case for measurements of properties other than position. Think, for instance, of a measurement of the momentum of an electron; where would such a quantum event be located? Clearly there is no sensible answer to such a question. This problem is a manifestation of the fact that, in the same way as standard quantum mechanics, RQM limits itself to make predictions for the possession of values of physical variables, but remains silent about what are the objects capable of possessing such values or, more generally, where are those properties supposed to be located. As such, it seems to perpetuate the ontological problems of the standard framework.

Proponents of RQM could argue that the postulation of property bearers for the values of properties is not necessary. That the acquisition of values themselves could constitute the ontology of the theory. In fact, \cite{Rov18} suggests that one should give up the notion of a primary substance, capable of carrying attributes, and to substitute it by ``the mutual dependence of the concepts we use to describe the world''. That is, that one must give primacy to `relations' over `objects'. While one could find such proposals suggestive or intriguing, it is not at all clear whether the idea of having ``relations without relata'' is even coherent. One could argue that the primacy of relations over objects simply means that all the (contingent) properties of objects are relative only. This, of course, would be cheating because, for this to work, one has to first postulate the existence of objects. What would be needed, of course, is a well-developed philosophical underpinning of this sort of ideas. In this regard, Rovelli, as well as works such as \cite{Dorato16,Candiotto17}, have argued that the best developed philosophical framework for this would be Ontic Structural Realism, which argues for the priority of relations over objects \citep{LyR}.

Another issue we would like to mention has to do with what could be called the ontological status of correlations. As it is clear from our overview of RQM, correlations play a very important role in RQM. However, given an ontological proposal solely based on quantum events, it is not clear what is meant when it is argued that a correlation obtains. Consider again a Wigner's friend scenario. According to RQM, a measurement of $S$ by $O$ produces a value actualization for $O$, but not for $P$. As a result, we are told, all there is for $P$ is a correlation between $O$ and $S$. However, if we take the quantum events ontology seriously, which indicates that all there is in the world are values actualizations, then we would have to conclude, not that what there is for $P$ is a correlation, but that, literally speaking, there is nothing. And, of course, without there being things to be correlated, the notion of a correlation is fully empty. % In other words, when we are told that, relative to some observer, a wave function is entangled, it is not at all clear to which (relative) fact is that supposed to correspond. Needless to say, this seems quite counter-intuitive and appears to lead to some sort of quasi-solipsistic world-view. 

%whether statements regarding the scope and virtues of RQM, in fact, hold. To see this, consider again a Wigner's friend scenario. According to RQM, a measurement of $S$ by $O$ produces a value actualization for $O$, but not for $P$. As a result, we are told, for $P$, all there is, is a superposition of the $O+S$ system, with terms corresponding to all the possible results of $O$'s measurement. However, if we take the quantum events-based ontology seriously, which indicates that all there is in the world are values actualizations, then we would have to conclude, not that what there is for $P$ is a superposition, but that, literally speaking, there is nothing. Needless to say, this seems quite counter-intuitive and appears to lead to some sort of quasi-solipsistic world-view. 

On top of all this, there is another serious problem for the proposed ontology. As we explained above, the idea is to define quantum events as the actualization of definite values. However, such a definition assumes that the time at which such actualizations occur is well-defined. The problem is that, as we explained in the previous section, this is not the case. As we saw, Rovelli seeks to define such a `time of interaction' in terms of the establishment of certain correlations, but the notion of a correlation, having an essential arbitrary component, is inadequate to support the definition. Consider, for instance, the measurement of any property of an electron with a measurement apparatus. Clearly, the interaction between the electron and the apparatus will be electromagnetic in nature, which implies it being ``turned-on'' at all times. How, then, could the time of the interaction be defined? Moreover, as we saw in Section \ref{M}, the establishment of correlations between, say, $S$ and $O$, is something that can be predicated only regarding the collective state of the $ S-O$ system which, within RQM, only exists for external observers $ P, P', P''$. What is more, for such a state to exist, there must be an interaction between the $ S-O$ system and one external observer, bringing us into a circular explanatory chain that seems clearly unacceptable as the foundation of a theory.

We conclude this section by pointing out that RQM heavily relies on the notion of an individual system (which can take a passive role, being subject to a characterization by another system, or an active one, characterizing other systems). Moreover, we are told that \emph{anything} can play such a role, without any restrictions of spatial or temporal extension. It seems, then, that highly non-local entities, such as, say, the electromagnetic field throughout space-time, could play the role of an individual system. This could lead to serious dangers, such as those described in \cite{Sorkin}. We also note that the notion of an interaction enters prominently into the proposal, but it is not clear what is the ontological status of such a concept. On one hand, RQM wants to be realist about interactions, but, on the other, if the world is only populated by quantum events, it is not clear that interactions can be part of the ontology. In any case, we will not dwell further on these issues in this paper.

Summing up, it is not only that an ontology based on quantum events is problematic, but that it is not even clear the quantum events themselves are well-defined. We conclude that RQM does not solve the ontological problems of the standard framework.

%%%%%%%%%%%%%%%%%%%%%%%%%%%%%%%%
\subsection{Consistency}
\label{C}
%%%%%%%%%%%%%%%%%%%%%%%%%%%%%%%%

As we mentioned above, the fully relational character of RQM might hint at potential inconsistencies. In particular, given that RQM fully embraces the notion that different observers may give different descriptions of the same sequence of events, it is natural to wonder whether RQM could lead to a paradoxical situation in which different observers get different results out of measurements of the same property of a given system. Moreover, even if self-consistent, such a state of affairs would seem hard to reconcile with the fact that descriptions made by different observers are (generally) found to be consistent. For instance, if a microbiologist describes a new virus with certain physicochemical properties, another scientist could be able to check the veracity of those observations in his own laboratory. If RQM asserts that every physical property is a relational property, associated with a particular observer, it is not clear how inconsistencies are avoided or why we do seem to experience observer-independent properties, at least at the macroscopic level.

In response to this type of worries, Rovelli warns that the issue must be handled with care. In particular, he highlights the fact that, within RQM, the information possessed by different observers cannot be compared or contrasted in the absence of physical interactions between them. For instance, in the context of a Wigner's friend scenario, we recall that, even if $O$ finds $q=1$, it is still possible that, when $P$ later interacts with $S-O$, she finds $q=2$. Isn't that inconsistent? The point is that, within RQM, what is meaningful to ask is whether there is an observer for which $O$ gets $q=1$ but $P$ gets $q=2$. In particular, regarding $P$, what is meaningful to ask is whether her measurement of $S$ coincides, not with what $O$ found when he measured $S$---as those measurements cannot be directly compared---but with what $O$ answers $P$ that he found when $P$ asks $O$ about his result.

Rovelli argues that quantum mechanics automatically guarantees the sort of consistency that RQM requires. This is because, after $O$ measures, $P$ assigns the $S-O$ system a superposition of $S$ measuring $q=1$ and $q=2$. Therefore, when $P$ measures $S$, she collapses the state of the full $S-O$ system, ensuring that the second measurement, i.e., asking $O$ what he got, yields a consistent answer.

In sum, RQM does not demand for measurements of the same system by different observers to always coincide---that would be an observer-independent requirement. Instead, RQM only demands for the sequence of events of each observer to be self-consistent. This, for Rovelli, is enough to ensure the empirical success of RQM, and to remove potential charges of solipsism against it.

Before moving on, it is important to point out that, in this whole explanation, which is all Rovelli offers on the matter, the question as to what $O$ \emph{experiences} (in case $O$ is a conscious being), in the event that he initially found $q=1$, but $P$ latter finds $q=2$, is never addressed. It is not clear whether $O$ remembers or not his previous result, as answers to these issues are ever offered.

In \cite{Rov07}, the issue of self-consistency of RQM is revisited in a more challenging context: an EPR experiment (in Bohm's formulation). They consider the standard scenario, in which a singlet is created, one particle of the pair is sent to $A$ and the other to $B$, and they measure them at space-like separated events. The reasoning in \cite{Rov07} is as follows. Since $A$ and $B$ are space-like separated, there cannot exist an observer with respect to which both of these outcomes are actual, and therefore it is meaningless to compare the results of $A$ and $B$. As a result, $A$'s and $B$'s results are fully independent---e.g., they could very well coincide.

However, when $B$'s particle is back into causal contact with $A$, she could go on and measure it. If so, as stated in \cite{Rov07}, quantum mechanics guarantees that, regardless of $B$'s result, $A$ will find the opposite of what $B$ found on his particle. Moreover, it is argued that, if $A$ goes on and asks $B$ what he found, quantum mechanics guarantees that, regardless of what $B$ found, $B$'s answer will be consistent with what $A$ found when she measured the particle. Finally, \cite{Rov07} claims that if a third observer asks both $A$ and $B$ for their results, perfect anti-correlation is guaranteed in the answers. The arguments for all this are analogous to the one discussed above regarding the Wigner's friend scenario: a measurement of $B$'s particle by $A$ collapses to a branch that guarantees the required correlations, and similarly for the third observer.

The lesson derived from all this is that, what RQM provides, is not the collection of all properties relative to all systems (such collection is assumed not to make sense), but one self-consistent description per system. Yet, systems can interact, and any system can be observed by another system. This, as stated in \cite{Rov07}, implies that any two observers can in fact communicate and, moreover, that either account of such interaction is guaranteed to be self-consistent. Inconsistencies arise only if one insists on an absolute state of affairs, obtained by juxtaposing descriptions relative to different observers. They conclude that RQM is the stipulation that this self-consistent, individual descriptions, is all one can talk about and, more importantly, that this is enough ``for describing nature and our own possibility of exchanging information about nature (hence circumventing solipsism)'' \cite{Rov07}. We, on the other hand, are not so sure about this last conclusion.

The point is that, in all scenarios considered above, there is an asymmetry between the observers involved, (e.g., $P$ is outside and measures after $O$; $A$ goes and measures $B$'s particle, and not the other way around), which is exploited to (allegedly) avoid trouble. However, when such an asymmetry is removed, the problems with the proposal become clear. To see this, consider again an EPR scenario, and suppose that an experiment is conducted as follows:
\begin{enumerate}
\item $A$ and $B$ receive a particle each and they measure it along the same axis, at space-like separation from each other; suppose that, in accordance with the possibilities allowed by RQM, they both obtain spin-up. 
\item After performing such a measurement, both of them travel to the mid-point between their labs and, at the same time, announce their results.
\end{enumerate}

What would happen then? There's a list of alternatives:
\begin{itemize}
\item One possibility is that, contrary to what RQM asserts, it is in fact metaphysically impossible for them to find the same result. This, of course, would mean that RQM is incomplete, which would be devastating for the proposal. 
\item Another option is that, for some reason, either $A$ or $B$ would announce the wrong result. This is exactly what is argued to happen in the asymmetric scenarios considered above, such as in Wigner's friend scenario, if $P$ asks $O$ about his result, in a case in which $O$ and $P$ find different values for $q$. However, in such a scenario, when $P$ measures $S$, she must interact with the whole $S-O$ system, so it is at least plausible that such an interaction could modify the state of $O$ in such a way that he ends up believing that he obtained a result for $q$ different from what he actually did (as we said above, RQM is mute on this issue). In the symmetric EPR scenario under consideration, on the other hand, an explanation of that sort seems out of place. To begin with, the situation is fully symmetric between $A$ and $B$, so it is not clear how it could be decided who would end up announcing the wrong result; and certainly, RQM does not have the means to do so. Moreover, unlike the Wigner's friend case, there does not seem to be any sort of mechanism that could explain why would, either $A$ or $B$, report a different result from what they actually obtained. Only stipulating that this is what would need to happen is, of course, not satisfactory.
\item Yet another option would be for both $A$ and $B$ to report either the right or the wrong result. That, of course, would preserve the symmetry, but would imply both reporting the same result, which would be empirically inadequate.
\item A final way out would be to allow for a total disconnect either between what each observer announce and what they think they announce, or between what they announce and the other hears. But that would clearly lead us deep into a solipsistic description.
\end{itemize}

From all this, it is clear that RQM has no sensible way to compare different descriptions of the same sequence of events, something we expect from any reasonable scientific theory. As we have said, RQM only establishes that an observer's view will be consistent with itself, but fully fails to establish any sort of relation between the descriptions provided by different observers.

In response to an EPR situation like this, in \cite{Rov07} is it argued that ``we shall not deal with properties of systems in the abstract, but only properties of systems relative to \emph{one} system. In particular, we can never juxtapose properties relative to different systems.'' It is then concluded that a case in which both $A$ and $B$ obtain spin-up ``can never happen either with respect to $A$ or with respect to $B$. The two sequences of events (the one with respect to $A$ and the one with respect to $B$ are distinct accounts of the same reality that cannot and should not be juxtaposed.'' We honestly fail so see how all this addresses the problem. As we showed above, RQM offers no way to compare different descriptions of the same sequence of events, something we expect that any scientific theory should do. It only provides for an observer's view to be consistent with itself, but fails to establish any relation between the descriptions offered by different observers. The specter of solipsism endures.

As we explained above, even if one were to grant that RQM is self-consistent, the program still has an important job to do. It has to explain how, out of a sea of observer-dependent descriptions, the apparently absolute collection of facts of the macroscopic world, emerges. \cite{Rov20} attempts to give an answer to such an issue employing the process of decoherence. The idea is that, out of a plethora of relative events, decoherence causes the emergence of \emph{stable} events or facts, whose relativity can effectively be ignored. Such stable facts explain, in turn, the emergence of the macroscopic, classical world.

\cite{Rov20} starts by pointing out that, in the classical, macroscopic world, if we know that one of a set of mutually exclusive facts $a_i$ has happened, then the probability for a fact $b$ to happen is given by
\begin{equation}
\label{sf}
P(b)= \sum_i P(b|a_i) P(a_i).
\end{equation}
\cite{Rov20} takes this equation as a characterization of \emph{stable facts}. In RQM, on the other hand, facts are of course relative. In particular, if an interaction between $F$ and $S$ affects $F$ in a manner that depends on the value of a certain variable $L_S$ of $S$, then the value of $L_S$ is a fact relative to $F$. In that case, the equation above has to be modified as such
\begin{equation}
\label{rf}
P(b^F)= \sum_i P(b^F|a_i^F) P(a_i^F),
\end{equation} 
where $a^F$ and $b^F$ are facts relative to $F$. Therefore, if one wants to explain within RQM the emergence of the macroscopic, classical world, then one has to show how, under appropriate circumstances, and at least approximately, one can write
\begin{equation} \label{stable}
P(b^W)= \sum_i P(b^W|a_i^F) P(a_i^F),
\end{equation} 
where $b^W$ are facts relative to an observer $W\neq F$.

How is this transition from relative to stable facts to be accomplished? In \cite{Rov20} it is claimed that the answer lies within decoherence. The idea is that such a phenomenon explains why (and when) we are allowed to say that, with respect to $W$, the state of the system $F$ ``has collapsed,'' without requiring an interaction between $W$ and $F$. In their words ``Decoherence clarifies why a large class of relative facts become stable with respect to us and form the stable classical world in which we live.''

The argument provided, goes as follows. Consider systems $F$, $W$ and the environment $\mathcal{E}$, and a variable $L_F$ of $F$; let $Fa_i$ be the eigenvalues of $L_F$. A generic state of the coupled system $F-\mathcal{E}$ can then be written as
\begin{equation}
\ket{\psi}=\sum_i c_i \ket{Fa_i}\otimes \ket{\psi_i},
\end{equation}
with $\ket{\psi_i}$ the states of $\mathcal{E}$. Next, assume: (a) that the states of the environment that correspond to different states of the system are almost orthogonal, and (b) that system $W$ does not interact with $\mathcal{E}$. Then, the probability $P(b)$ of any possible fact relative to $W$, resulting from an interaction between $F$ and $W$, can be computed from the density matrix obtained by tracing over $\mathcal{E}$, that is
\begin{equation} \label{reduced}
\rho=Tr_{\mathcal{E}}\ket{\psi}\bra{\psi}=\sum_i |c_i|^2 \ket{Fa_i}\bra{Fa_i} + O(\epsilon),
\end{equation} 
where $\epsilon=max_{i,j}|\braket{\psi_i|\psi_j}|^2$. Finally, by posing $P(Fa_i^{\mathcal{E}})=|c_i|^2$, the authors write
\begin{equation}
P(b^W)=\sum_i P(b^W|Fa_i^{\mathcal{E}})P(Fa_i^{\mathcal{E}}) + O(\epsilon).
\end{equation} 
Thus, they claim, to the extent to which one ignores effects of order $\epsilon$, the facts $L_F=Fa_i$ relative to $\mathcal{E}$ become stable with respect to $W$.

Unfortunately, as we already pointed out, decoherence by itself is unable to explain the emergence of the classical (stable) world around us, and it also fails in this context. The problem here, as in other arguments involving decoherence, is that the reduced density matrix in (\ref{reduced}) is interpreted, \textit{for all practical purposes}, as a classical mixture. However, in order to show that the decohered system indeed behaves like the mixture, one needs to assume the Born rule between $\mathcal{E}$ and $F$. That is, $W$ needs to assume that the environment measures $F$ and gets a single outcome. But this is precisely what is forbidden by RQM: $W$ cannot make predictions assuming that a collapse already took place between $\mathcal{E}$ and $F$---this is only possible for stable facts. Therefore, in order to show that stable facts will emerge, one needs to precisely assume stable facts beforehand. We conclude that the argument provided in \cite{Rov20} to explain the emergence of the classical (stable) world in RQM begs the question. 

Rovelli often claims that the transition from standard quantum mechanics to RQM is analogous to the construction of special relativity by Einstein in 1905. In both cases, he argues, the obstruction to a correct understanding of a preexisting theoretical formalism was removed by the recognition that an element previously thought to be absolute is in fact relative (``simultaneity'' in special relativity and ``states'' in the case of RQM). Moreover, in \cite{Rov07} it is claimed that ``the meaning of the adjective ``relative'' in the RQM notion of ``relative state'' is... very similar to the meaning of ``relative'' in special relativity.'' We point out, though, that there is an important aspect in which the analogy breaks down. In the setting of special relativity, it is clear that the descriptions associated with different observers can always be combined, and derived from, a single unified description provided by a geometric object, such as Minkowski's space-time, the world-lines within it, etc. This, of course, guarantees coherence and consistency when different observers interact or compare notes. RQM, on the other hand, explicitly lacks such a single unified description, i.e., it does not contain a mathematical object encoding all the different perspectives. The problem, as we just saw, is that without it, RQM cannot really accommodate a coherent and consistent interchange of information between observers. That is, after all, RQM seems not to contain enough resources to adequately explain the possibility of exchanging information about the world between observers, or to explain the emergence of an apparently absolute, observer-independent macroscopic world. 

%%%%%%%%%%%%%%%%%%%%%%%%%%%%%%%%
\subsection{Locality}
\label{L}
%%%%%%%%%%%%%%%%%%%%%%%%%%%%%%%%

The status of locality within RQM has been discussed mainly in two works, \cite{Rov07} and \cite{Rov19}. In the first one, it is argued that if one acknowledges that, in the absence of a physical interaction between observers, it is meaningless to compare measurements---and thus, that it is meaningless to compare measurements occurring at space-like separation---then one concludes that the EPR correlations do not entail any form of non-locality. From that, they conclude that RQM is a fully local theory. However, in \cite{Rov19}, the claim that RQM is fully local is retracted, and the fact that RQM is non-local in Bell's sense is granted. Yet, it is argued that, in the context of RQM, the failure of locality is not deep or interesting because it simply ``reduces to the existence of a common cause in an indeterministic context''. In this section we will explore in detail all these claims. However, before doing so, first we need to properly construe the notion of locality.

In order to define locality in a precise way, we, of course, follow Bell. Bell's definition of locality relies on the notion of \emph{local beables}. The beables of a theory are defined by Bell to be whatever is posited by the theory to correspond to things that are physically real. Local beables, in turn, are beables which are definitely associated with particular space-time regions. It is clear that what Bell has in mind when he talks about ``something being real'' is for it to have an \emph{observer-independent} reality. Therefore, strictly speaking, there are no beables within RQM. Still, in case the issues we discussed in section \ref{O} could be addressed, RQM could be argued to have \emph{beables relative to an observer}. As for the locality of such observer-dependent beables, as we saw, it is quite obscure, in general, where in space-time they would be located. 

To define locality in terms of local beables, Bell introduces what he calls the \emph{principle of local causality} (see, e.g., \cite{Bell1990}). According to such a principle, a theory is \emph{local} if the probability it assigns to $b_\chi$, the value of some beable at the space-time event $\chi$, is such that
\begin{equation} \label{condprobA}
P(b_\chi|\lambda_\sigma) = P(b_\chi| \lambda_\sigma,b_\xi),
\end{equation} 
with $\lambda_\sigma$ a complete specification of the physical state on $\sigma$, a spatial slice fully covering the past light cone of $\chi$, and $b_\xi$ the value of any beable or property on an event $\xi$, space-like separated from $\chi$ and outside of the causal future of $\sigma$ (see Figure 1).
%%%%%%%%%%%%%%%%%%%%%%%%%%%%%%
\begin{figure}[ht]
\centering
\includegraphics[height=6cm]{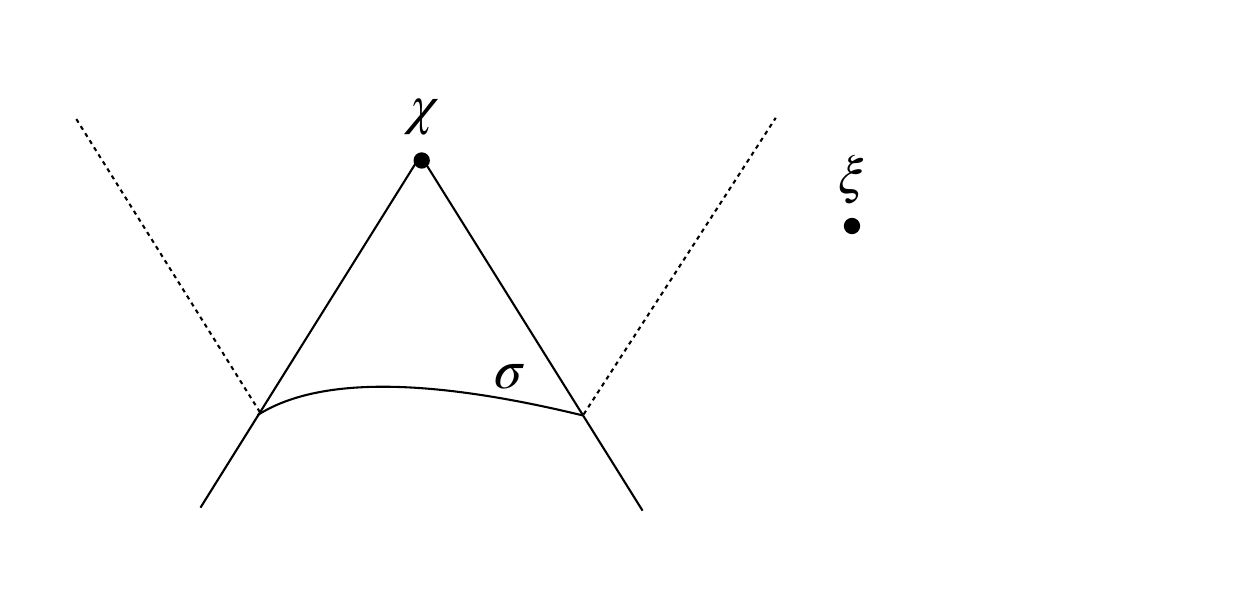} 
\caption{According to Bell's principle of local causality, a theory is local if $P(b_\chi|\lambda_\sigma) = P(b_\chi|\lambda_\sigma,b_\xi)$.}
\end{figure}
%%%%%%%%%%%%%%%%%%%%%%%%%%%%%%
That is, for local theories, if complete information on a slice of the past light cone of an event is available, then new information regarding happenings on regions outside of the future of such a slice cannot alter the predictions of the theory regarding that event. Note that Bell's definition applies equally well to deterministic and indeterministic theories. Note also that it is crucial for $\xi$ to lie outside of the causal future of $\sigma$. Otherwise, even for a local theory, $\xi$ could provide information about $\chi$. That is because, in an indeterministic theory, a stochastic process in the future of $\sigma$, but in the common past of $\chi$ and $\xi$, could correlate them (see \cite[p.16]{norsen2017} for details).

As is well-known, based on the definition of locality described above, Bell showed that no local theory is able to correctly account for all quantum correlations, \cite{Bell1964,Bell1971,Bell1976,Bell1990}. That is, he showed that quantum correlations cannot be explained via a local common cause.

As it is clear from the definition, Bell's locality not only depends on local beables (e.g., it assumes there is such a thing as $b_\chi$, a beable at $\chi$, or $\lambda_\sigma$, the complete state on region $\sigma$), but presupposes all beables to be local. In fact, it seems difficult, if not impossible, to construct a notion of locality without assuming the existence of at least some local beables---and it is not even clear what role could non-local beables play in a local theory. In any case, in order to declare that a theory without local beables is local, it is necessary to show that it is indeed possible to define locality in the absence of local beables, and to clearly state what alternative to Bell's definition is being employed. 

The proposal in \cite{Rov07}, in this regard, basically amounts to substituting Bell's beables with RQM's relational beables. Granting for the moment that the latter are indeed well-defined and local, such a strategy would seem to lead to a reasonable proposal. The concrete reformulation of locality offered in \cite{Rov07} reads: \emph{relative to a given observer, two spatially separated events cannot have instantaneous mutual influence.} With that definition in hand, in the context of an EPR experiment, it is then argued that, since $A$ and $B$ are space-like separated, there cannot exist an observer with respect to which their results are actual, so it is meaningless to compare them. From this, it is argued that, in the context of RQM, it is not necessary to abandon locality in order to account for EPR correlations. This, in turn, is taken to imply that RQM is a fully local theory.

However, as straightforwardly recognized in \cite{Rov19}, even within RQM, there are plenty of observers for which the results of $A$ and $B$ are in fact actual: any observer that lies in the common future of $A$ and $B$ would do. Still, in \cite{Rov19} it is argued that those future observers would simply conclude that there is a common cause to $A$ and $B$---albeit a indeterministic one. From this, they conclude that there is no need to invoke mysterious space-like influences to understand the correlations. That is, that within RQM, the weirdness of non-locality reduces to the weirdness of indeterminism. In other words, it is claimed that Bell's definition of locality fails to correctly capture such a notion in an indeterministic context.

All this, of course, is simply incorrect. As we mentioned above, Bell's definition of locality is perfectly suitable for indeterministic contexts. Moreover, as we also mentioned above, Bell's work precisely proves that the correlations between $A$'s and $B$'s result, as experienced by an observer in their common future, cannot be explained by the postulation of a local common cause, be it deterministic or indeterministic. We conclude that the conclusion in \cite{Rov19} to the effect that, within RQM, mysterious space-like influences are not required to explain observed correlations is simply false. That is, even within the terms proposed in \cite{Rov07} and \cite{Rov19}, and to the extent that one grants RQM's relational beables to be well-defined and local, RQM ends up being fully non-local.

%%%%%%%%%%%%%%%%%%%%%%%%%%%%%%%%%%%%%%%%%%%%%%%%%%%%%%%%%%
%%%%%%%%%%%%%%%%%%%%%%%%%%%%%%%%%%%%%%%%%%%%%%%%%%%%%%%%%%
\section{Conclusions}
\label{Conc}
%%%%%%%%%%%%%%%%%%%%%%%%%%%%%%%%%%%%%%%%%%%%%%%%%%%%%%%%%%
%%%%%%%%%%%%%%%%%%%%%%%%%%%%%%%%%%%%%%%%%%%%%%%%%%%%%%%%%%

RQM proposes to get rid of the notion of absolute states of systems, in favor of states of systems in relation to each other. Such a move is claimed to solve the well-known conceptual problems of standard quantum mechanics. Moreover, in spite of adopting a purely relational character, RQM has been argued to fully and successfully describe our experience of the world and to be able to explain how observers exchange information. Finally, RQM has been claimed to account for all quantum correlations, without the need of invoking any sort of non-local influences.

In this work, we carried out a thorough assessment of RQM. Regarding the claim that RQM solves the conceptual problems of standard quantum mechanics, we found that it fails to directly address such issues, and that it leads to serious conceptual problems of its own. On one hand, it perpetuates the vagueness issues of the standard interpretation and, on the other, does nothing to alleviate its lack of ontological clarity.

As for the beliefs that RQM can correctly explain our experience of the world, including our interactions with other observers, and that it accommodates all quantum correlations, without invoking non-local effects, we found them to be unwarranted. RQM is an original and creative attempt to deal with the conceptual difficulties of quantum theory. Unfortunately, it does not fare well under careful scrutiny. We conclude that RQM is unable to provide a satisfactory understanding of the quantum world.\footnote{After posting this manuscript online, some of the issues with RQM described in it have generated a very fruitful discussion (for the preferred-basis problem see \cite{brukner2021qubits,pienaar2021quintet,stacey2021relational} and for the exchange of information between observers see \cite{adlam2022does,adlam2022information}).}

%%%%%%%%%%%%%%%%%%%%%%%%%%%%%%%%%%%%%%%%%%%%%%%%%%%%%%%%%%
%%%%%%%%%%%%%%%%%%%%%%%%%%%%%%%%%%%%%%%%%%%%%%%%%%%%%%%%%%
\section*{Acknowledgments}
%%%%%%%%%%%%%%%%%%%%%%%%%%%%%%%%%%%%%%%%%%%%%%%%%%%%%%%%%%
%%%%%%%%%%%%%%%%%%%%%%%%%%%%%%%%%%%%%%%%%%%%%%%%%%%%%%%%%%
We would like to thank Quentin Ruyant, Tim Maudlin and Travis Norsen for valuable comments. We acknowledge partial financial support from PAPIIT-DGAPA-UNAM project IG100120 and CONACyT project 140630. DS is grateful for the support provided by the grant FQXI-MGA-1920 from the Foundational Questions Institute and the Fetzer Franklin Fund, a donor advised by the Silicon Valley Community Foundation.
%%%%%%%%%%%%%%%%%%%%%%%%%%%%%%%%%%%%%%%%%%%%%%%%%%%%%%%%%%
%%%%%%%%%%%%%%%%%%%%%%%%%%%%%%%%%%%%%%%%%%%%%%%%%%%%%%%%%%
%\bibliographystyle{plain}
\bibliographystyle{apalike}
\bibliography{bibRQM.bib}
%%%%%%%%%%%%%%%%%%%%%%%%%%%%%%%%%%%%%%%%%%%%%%%%%%%%%%%%%%
%%%%%%%%%%%%%%%%%%%%%%%%%%%%%%%%%%%%%%%%%%%%%%%%%%%%%%%%%% 

\end{document}